# When the map is better than the territory

Erik P Hoel*

*Department of Biological Sciences, Columbia University, New York, NY, USA*
*hoelerik@gmail.com*

**Abstract**. The causal structure of any system can be analyzed at a multitude of spatial and temporal scales. It has long been thought that while higher scale (macro) descriptions of causal structure may be useful to observers, they are at best a compressed description and at worse leave out critical information. However, recent research applying information theory to causal analysis has shown that the causal structure of some systems can actually come into focus (be more informative) at a macroscale (Hoel et al. 2013). That is, a macro model of a system (a map) can be more informative than a fully detailed model of the system (the territory). This has been called "causal emergence." While causal emergence may at first glance seem counterintuitive, this paper grounds the phenomenon in a classic concept from information theory: Shannon's discovery of the channel capacity. I argue that systems have a particular causal capacity, and that different causal models of those systems take advantage of that capacity to various degrees. For some systems, only macroscale causal models use the full causal capacity. Such macroscale causal models can either be coarse-grains, or may leave variables and states out of the model (exogenous) in various ways, which can improve the model's efficacy and its informativeness via the same mathematical principles of how error-correcting codes take advantage of an information channel's capacity. As model choice increase, the causal capacity of a system approaches the channel capacity. Ultimately, this provides a general framework for understanding how the causal structure of some systems cannot be fully captured by even the most detailed microscopic model.

**Introduction**. It has been difficult to find a causal role for the macroscales of a system (Fodor 1974; Kim 2000). But a way forward is to consider this issue not as a philosophical problem but rather as a problem of causal model choice. Causal models are those that represent the influence of subparts of a system on other subparts, or over the system as a whole. A causal model may represent state transitions, like Markov chains, or may represent the influence or connectivity of elements, such as circuit diagrams, directed graphs (also called causal Bayesian networks), networks of interconnected mechanisms, or neuron diagrams. Using causal models in the form of networks of logic gates, actual *causal emergence* was first demonstrated in Hoel et al. (2013). Causal emergence is when the macro beats the micro in terms of the efficacy, informativeness, or power of its causal relationships. It is identified by comparing macroscales to their underlying microscales, and analyzing the causal structure of both using information theory. Here it is revealed that causal emergence is related to a classic concept in information theory, Shannon's channel capacity (Shannon 1949), thus grounding it rigorously in another well-known mathematical phenomenon.

There is a natural, but unremarked upon, connection between causation and information theory. Both causation and information are defined in respect to the nonexistent: causation relies on counterfactual statements, and information theory on unsent signals. While their similarities have not been enumerated explicitly, it is unsurprising that a combination of the two has begun to gain traction. Causal analysis is performed on a set of interacting elements or state transitions, i.e., a causal model, and the hope is that information theory is the best way to quantify and formalize these interactions and transitions.

Some information-theoretic constructs already use quasi-causal language; measurements like granger causality (Granger 1969), directed information (Massey 1990), and transfer entropy (Schreiber 2000). Despite being interesting and useful measures, these either fail to accurately measure causation, or are too vague and underdeveloped (Janzing et al. 2013). More explicit attempts to tie information theory to causation use Judea Pearl's causal calculus (Pearl 2000). These measures rely on intervention, formalized as the $do(x)$ operator (the distinctive aspect of causal analysis). Stated in the most general manner possible, the idea is to calculate the mutual information $I(X,Y)$ between a set of interventions ($X$) and their effects ($Y$). One such measure is the information flow (Ay and Polani 2008), which was essentially renamed and rediscovered in Korb et al. (2009). But the most explicit connection between causation and information can be traced back to a measure named *effective information* (*EI*), which assesses the causal influence of one subset of a system on another (Tononi 2001; Tononi and Sporns 2003). When *EI* is assessed over the entire system (that is, the mutual information between the full repertoire of system states) it captures how effective and informative a system's causal structure is. Used this way, *EI* represents a quantification of "deep understanding", defined by Judea Pearl as "knowing not merely how things behaved yesterday but also how things will behave under new hypothetical circumstances" (Pearl 2000). For instance, in Hoel et al. (2013) it was shown that *EI* is the average effect information of a system's states. Effect information is the causal influence of an individual system state, defined as how much the state reduces the uncertainty about the future of the system. *EI* is used here because of its previous usage to demonstrate causal emergence (Hoel et al. 2013), its relative historical precedence, its proven link to important properties of causal structure (Hoel et al. 2013), and mutual information's understandability and its role as the centerpiece of information theory. First *EI* is introduced, its relationship to causal structure is shown, its relationship to mutual information is proven, and then the measure is used to demonstrate how systems have a particular causal capacity that resembles an information channel's capacity, which different causal models of the system take advantage of to differing degrees.

**Assessing causal structure with information theory.** To measure *EI*, first the $do(x)$ operator is used to set a system $S$ into a particular state $s_i$ at some time $t$ and observing the effects at some time $t_{+1}$. More broadly, there is an application of some *Intervention Distribution* ($I_D$) composed of the probabilities of each $do(s_i)$. The results of applying $I_D$ to the system is the *Effect Distribution (E_D)*, that is, the effects of $I_D$. A uniform distribution of interventions is applied over the full set of system states such that in $I_D$ each $p$ is $1/n$, where $n$ is the number of states. This is identical to intervening upon the system using the maximum entropy distribution ($I_D = H^{max}$): perturbing some system $S$ into all $n$ possible initial states with equal probability ($1/n$) so that $(do(S = s_i) \forall_i \in 1 \ldots n)$. In a memoryless system with the Markov property $I_D$ is applied at some time $t$, and $E_D$ is the distribution of states transitioned into at $t_{+1}$.

From the application of $I_D$ a Transition Probability Matrix (TPM) is constructed using Bayes' rule. The TPM associates each state $s_i$ in $S$ with a probability distribution of past states ($S_P | S = s_i$) that could have lead to it, and a probability distribution of future states that it could lead to ($S_F | S = s_i$). Note that a TPM can be obtained for a system with specified elements and mechanisms, like a system of interconnected logic gates, or for a Markov chain defined only as state transitions, or from a Bayesian causal network, and so on. In fact, the defining feature of a causal model seems to be that it can be represented as a TPM.

*EI* quantifies the architecture of the TPM. In a discrete finite system, *EI* is:

$$EI(S) = \frac{1}{n} \sum_{s_i \in S} D_{KL} \left(S_F | do(S = s_i) || E_D\right)$$

where $n$ is the number of system states and $D_{KL}$ is the Kullback-Leibler divergence (Kullback 1997). The effect information of an individual state is $D_{KL}(S_F | do(S = s_i) || E_D)$, which is the difference that state $s_i$ makes to the future of the system. If all states transition with equal probability to all states then no state makes any difference to the future of the system. Conversely, if all states transition to the same state, then no state makes any difference to the future of the system. If each state transitions to a single other state that no other state transitions to, then and only then does each state make the maximal difference to the future of the system. Only then is *EI* maximal.

The relationship between causation and information was directly proven in Hoel et al. (2013). There it was shown that *EI* is equal to the mutual information $I(C,E)$, where $C$ is a set of causes (defined as some set of states) and $E$ the set of their subsequent effects (the states they transition to). An identical formulation is as the mutual information between interventions and their effects, such that $EI = I(I_D, E_D)$. There it was also proven that $EI$ is sensitive to important properties of causal relationships like *determinism* (the noise of state-to-state transitions, which is also the sufficiency of a cause to accomplish an effect), *degeneracy* (how often states transition to the same state, which is also the necessity of a cause), and complexity (the size of the state-space, or the number of possible causes).

To demonstrate $EI$'s ability to accurately quantify causal structure, consider the TPMs ($t$ by $t_{+1}$) of three Markov chains, each with $n = 4$ states [00, 01, 10, 11]:

$$M_1 = \begin{bmatrix} 0 & 0 & 1 & 0 \\ 1 & 0 & 0 & 0 \\ 0 & 0 & 0 & 1 \\ 0 & 1 & 0 & 0 \end{bmatrix}.$$

$$M_2 = \begin{bmatrix} 1/3 & 1/3 & 1/3 & 0 \\ 1/3 & 1/3 & 1/3 & 0 \\ 0 & 0 & 0 & 1 \\ 0 & 0 & 0 & 1 \end{bmatrix}.$$

$$M_3 = \begin{bmatrix} 1/4 & 1/4 & 1/4 & 1/4 \\ 1/4 & 1/4 & 1/4 & 1/4 \\ 1/4 & 1/4 & 1/4 & 1/4 \\ 1/4 & 1/4 & 1/4 & 1/4 \end{bmatrix}.$$

Note that in $M_1$ every state completely constrains both the past and future, while the states of $M_2$ constrain the past/future only to some degree, and finally $M_3$ is entirely unconstrained (the probability of any state-to-state transition is $1/n$). This affects the chains' respective $EI$ values: $EI(M_1) = 2$ bits, $EI(M_2) = 1$ bit, $EI(M_3) = 0$ bits. Given that the systems are the same size ($n$), their differences in $EI$ stem from their differing levels of *effectiveness* (Fig. 1), a value which is bounded between 0 and 1. Effectiveness (*eff*) is how successful a causal structure is in turning interventions into unique effects. In Figure 1 the three chains are drawn above their TPMs. Probabilities are represented in grayscale (black is $p = 1$).

The effectiveness of a particular TPM is decomposable into two factors: the determinism and degeneracy of the TPM. Determinism is a measure of how reliable state-to-state transitions are, assessed by comparing the possible effects of each state ($S_F | S = s_i$) to the maximum entropy distribution $H^{max}$:

$$determinism(S) = \frac{1}{n} \sum_{s_i \in S} \frac{D_{KL}(S_F | do(S = s_i) || H^{max})}{\log_2(n)}$$

A low determinism is a mark of noise. Degeneracy on the other hand measures how much deterministic convergence there is among the effects of states (convergence/overlap not due to noise):

$$degeneracy(S) = \frac{D_{KL}(E_D || H^{max})}{\log_2(n)}$$

A high degeneracy is a mark of attractor dynamics. Together, determinism and degeneracy contribute equally to *eff*, which is how effective a system is at transforming interventions into effects, or past states into future ones:

$$effectiveness(S) = determinism(S) - degeneracy(S).$$

If a system has *eff* = 1 then all of its state-to-state transitions are logical biconditionals of the form $s_i \Leftrightarrow s_k$. A relationship which is a logical biconditional can only occur if $s_i$ is both completely sufficient and utterly necessary to produce $s_k$. See Hoel et al. (2013) for further details on effectiveness, determinism, and degeneracy. Additionally, see Hoel et al. (2016) for how *eff* contributes to how a causal relationship acts when partitioned.

The total $EI$ of a system depends not solely on *eff*, but also on the number of states. If the size of the system is $\log_2(n)$ of the number of system states, then:

$$EI(S) = size(S) * eff(S).$$

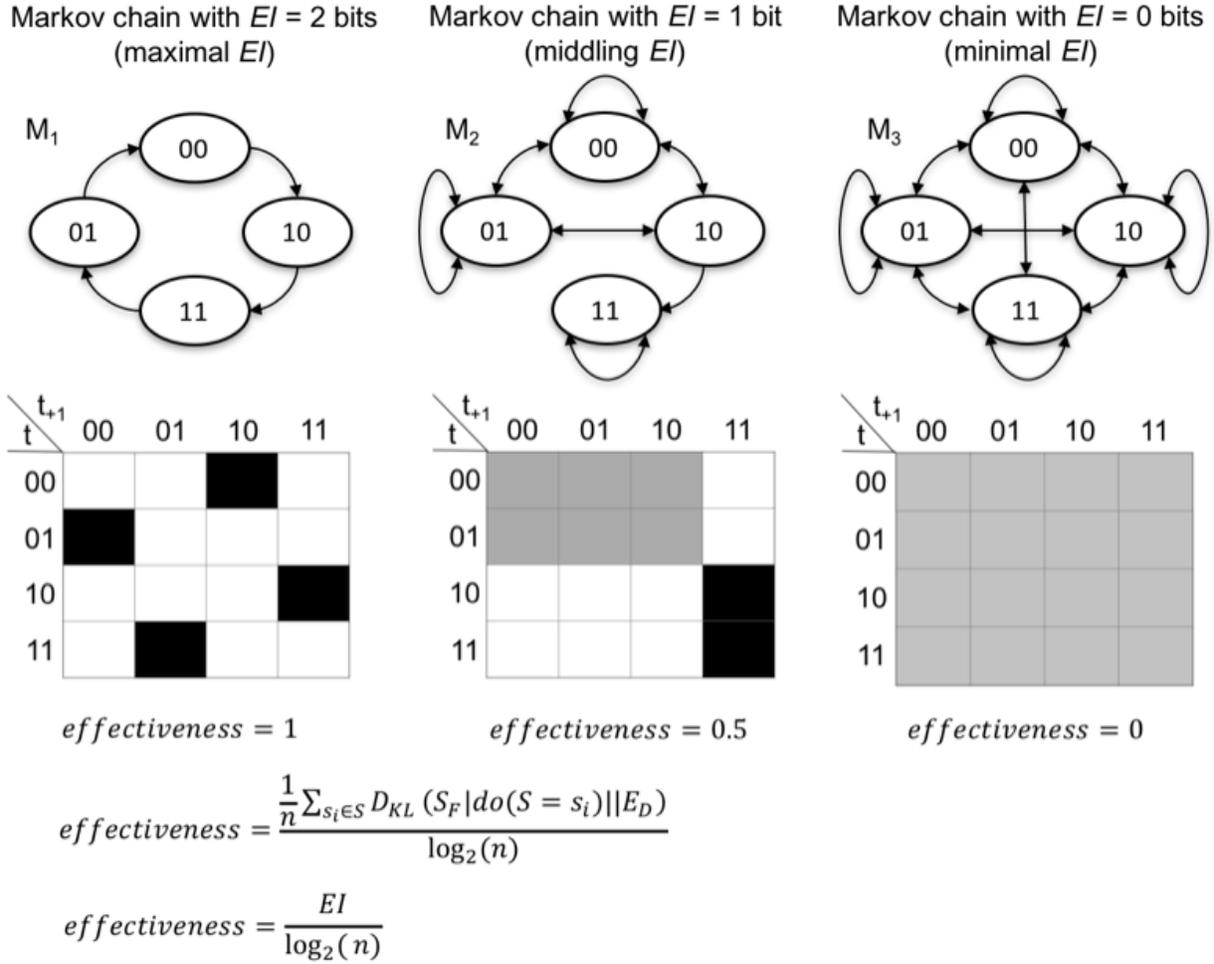

**Figure 1**. *Markov chains with different levels of effectiveness*

If two systems have equal *eff*, but one has a much larger *size*, then that system has greater *EI*. There is far more information in a causal model with a thousand interacting variables than two. For example, consider the causal relationship between a light switch and light bulb {LS, LB}. The effectiveness of the causal relationship is *eff* = 1 (as the on/off state of LS at *t* perfectly constrains the on/off state of LB at $t_{+1}$). Compare that to a light dial (LD) with 256 discriminable states, which controls a different light bulb (LB₂) that possesses 256 discriminable states of luminosity. Just examining the sufficiency or necessity of the causal relationships would not inform us of the crucial causal difference between these two systems. But *EI* would be 1 bit for $LS \rightarrow LB$ and 8 bits for $LD \rightarrow LB_2$, capturing the fact that a causal model with the same general structure but more states should contain more information, as well as that the causal influence of the light dial is correspondingly that much greater.

A system can have a smaller *size*, but as long as it has more *eff*, its *EI* can be higher. Consider two Markov chains: $M_A$ has a high *eff*, but low *size*, while $M_B$ has a large *size*, but low *eff*. If $eff_A > eff_B$, and $size_A < size_B$, then $EI(M_A) > EI(M_B)$ only if $\left(\frac{eff_B}{eff_A} > \frac{size_A}{size_B}\right)$. This means that if $size_B >> size_A$ then there must larger relative differences in effectiveness, such that $eff_A >>> eff_B$, for $M_A$ to have higher *EI*.

*EI* is a good measure of causal structure because it is sensitive solely to the structure of the TPM. As $I_D = H^{max}$, the marginal probability $p(x)$ is screened off and the analysis is solely dependent on the conditional probabilities $p(y|x)$, i.e., the causal structure. Notably, *EI* ultimately succeeds for the same reason that causal effects should be identified by the application of randomized trials (Fisher 1935).

**Causal analysis across scales**. Any system can be represented in different ways, via different causal models. Here, we only consider discrete systems with a finite number of states and/or elements. The microscopic causal model of such a system is its most fine-grained representation in space and time ($S_m$). But systems can also be considered at scales above $S_m$, as many different macro causal models ($S_M$). The set of all possible causal models, $\{\mathbf{S}\}$, is entirely fixed by the base $S_m$. In technical terms this known as supervenience: given the lowest scale of any system (the base), all the subsequent higher scales of that system are fixed (Davidson 1970; Stalnaker 1996). Due to multiple realizability, different $S_m$'s may share the same $S_M$.

Macro causal models are defined as a mapping: $M: S_m \rightarrow S_M$. A macroscale can be a mapping in space, time, or both. As they are similar mathematically, here we only examine spatial mappings (but see Hoel et al. 2013;2016 for temporal examples). One universal definitional feature of a macroscale, in space and/or time, is its reduced size: $S_M$ must always be of a smaller cardinality than $S_m$. For instance, a macro causal model may be a mapping of states that leaves out (considers exogenous) some of the states in the micro causal model.

Some macro causal models are coarse-grains: they map many microstates onto a single macrostate. Microstates, as long as they are mapped to the same macrostate, are treated identically at the macroscale (but note that they themselves don't have to be identical at the microscale). A coarse-grain over elements would group two or more micro-elements together into a single macro-element. Macrostates and elements are therefore similar to those in thermodynamics: defined as invariant of underlying micro-identities, whatever those might be. For instance, if two micro elements A & B are mapped into a macro element, switching the state of A and B should not change the macrostate. In thermodynamics, temperature is such a macrostate of molecular movement, and in neuroscience, an action potential is such a macrostate of ion channel openings. In terms of causal analysis, this means that a coarse-grained intervention is:

$$do(S_M = s_M) = \frac{1}{n} \sum_{s_{m,i} \in s_M} do(s_m = s_{m,i})$$

where $n$ is the number of microstates ($s_i$) mapped into $S_M$. Put simply, a coarse-grained intervention is an average over a set of micro interventions. In general, when a set of macro interventions are applied each member of $I_D$ is $1/m$ (where $m$ is the number of macrostates). That is, for the states explicitly represented in the macro model, $I_D$ is still applied equiprobably (and therefore accordance with Fisher (1935)). So despite the fact that $I_D$ may no longer equal $H^{max}$ over all (micro) system states, it will over the set of macrostates explicitly included in the macro causal model.

**Causal emergence.** A simple example of causal emergence is a Markov chain $S_m$ with $n = 8$ possible states, with the TPM:

$$S_m = \begin{bmatrix} 1/7 & 1/7 & 1/7 & 1/7 & 1/7 & 1/7 & 1/7 & 0 \\ 1/7 & 1/7 & 1/7 & 1/7 & 1/7 & 1/7 & 1/7 & 0 \\ 1/7 & 1/7 & 1/7 & 1/7 & 1/7 & 1/7 & 1/7 & 0 \\ 1/7 & 1/7 & 1/7 & 1/7 & 1/7 & 1/7 & 1/7 & 0 \\ 1/7 & 1/7 & 1/7 & 1/7 & 1/7 & 1/7 & 1/7 & 0 \\ 1/7 & 1/7 & 1/7 & 1/7 & 1/7 & 1/7 & 1/7 & 0 \\ 1/7 & 1/7 & 1/7 & 1/7 & 1/7 & 1/7 & 1/7 & 0 \\ 0 & 0 & 0 & 0 & 0 & 0 & 0 & 1 \end{bmatrix}$$

The effectiveness of $S_m$ is very low (*eff* = 0.18) and so $EI(S_m)$ is only 0.55 bits. A search over all possible mappings reveals a TPM of the macro causal model with $EI^{max}(S_M) = 1$ bit:

$$S_M = \begin{bmatrix} 1 & 0 \\ 0 & 1 \end{bmatrix}$$

demonstrating the causal emergence of 0.45 bits. In this mapping, the first 7 states have all been grouped into a single macrostate. Note that even though these microstates have been grouped into a macrostate this doesn't mean that these states actually have to be equivalent for there to be causal emergence. For instance, if the TPM of $S_m$ is:

$$S_m = \begin{bmatrix} 1/5 & 1/5 & 1/5 & 1/5 & 1/5 & 0 & 0 & 0 \\ 1/7 & 3/7 & 1/7 & 0 & 1/7 & 0 & 1/7 & 0 \\ 0 & 1/6 & 1/6 & 1/6 & 1/6 & 1/6 & 1/6 & 0 \\ 1/7 & 0 & 1/7 & 1/7 & 1/7 & 1/7 & 2/7 & 0 \\ 1/9 & 2/9 & 2/9 & 1/9 & 0 & 2/9 & 1/9 & 0 \\ 1/7 & 1/7 & 1/7 & 1/7 & 1/7 & 1/7 & 1/7 & 0 \\ 1/6 & 1/6 & 0 & 1/6 & 1/6 & 1/6 & 1/6 & 0 \\ 0 & 0 & 0 & 0 & 0 & 0 & 0 & 1 \end{bmatrix}$$

then the transition profiles ($S_F \,|\, S = s_i$) are different for each microstate. However, $EI^{max}$ is still at the same macro level ($EI(S_M) = 1$ bit $> EI(S_m) = 0.81$ bits).

How is it possible for the macro causal model to possess $EI^{max}$? While all macro causal models inherently have a smaller *size*, there may be an increase in *eff*. As stated previously, for two Markov chains, $EI(M_x) > EI(M_y)$ if $\left(\frac{eff_y}{eff_x} > \frac{size_x}{size_y}\right)$. Since the ratio of *eff* can increase to a greater degree than the accompanying decrease in the ratio of *size*, the macro can beat the micro.

Consider a generalized case of causal emergence: a Markov chain for which every $n - 1$ state has the same $1/n - 1$ probability to transition to any of the set of $n - 1$ states. The remaining state $n_z$ transitions to itself with $p = 1$. For such a system $EI^{max}(S_M) = 1$ bit, no matter how large the *size* of the system is (from a mapping $M$ of $n_z$ into macrostate 1 and all remaining $n - 1$ states into a macrostate 2). In this case as the *size* increases $EI(S_m)$ decreases: $\lim_{n \to \infty} EI(S_m) = 0$ as $\lim_{n \to \infty} 1/(n-1) = p = 0$. A macro causal model $S_M$ can remain the same even as the underlying microscale drops off to an infinitesimal $EI$. This also means that the upper limit of the difference between $EI(S_M)$ and $EI(S_m)$ (the amount of causal emergence) is theoretically bounded only by $\log_2(m)$, where $m$ is the number of macrostates.

Causal emergence is possible in completely deterministic systems, as long as they are degenerate. For instance, consider a network of binary logic gates, first introduced in Hoel et al. (2013). At $S_m$ it is six micro elements, all of which operate as deterministic AND gates (binary nodes which equal 1 at $t$ iff both inputs are 1 at $t_{-1}$). The connectivity and TPMs of both $S_m$ and the macroscale with $EI^{max}$ are shown in Figure 2.

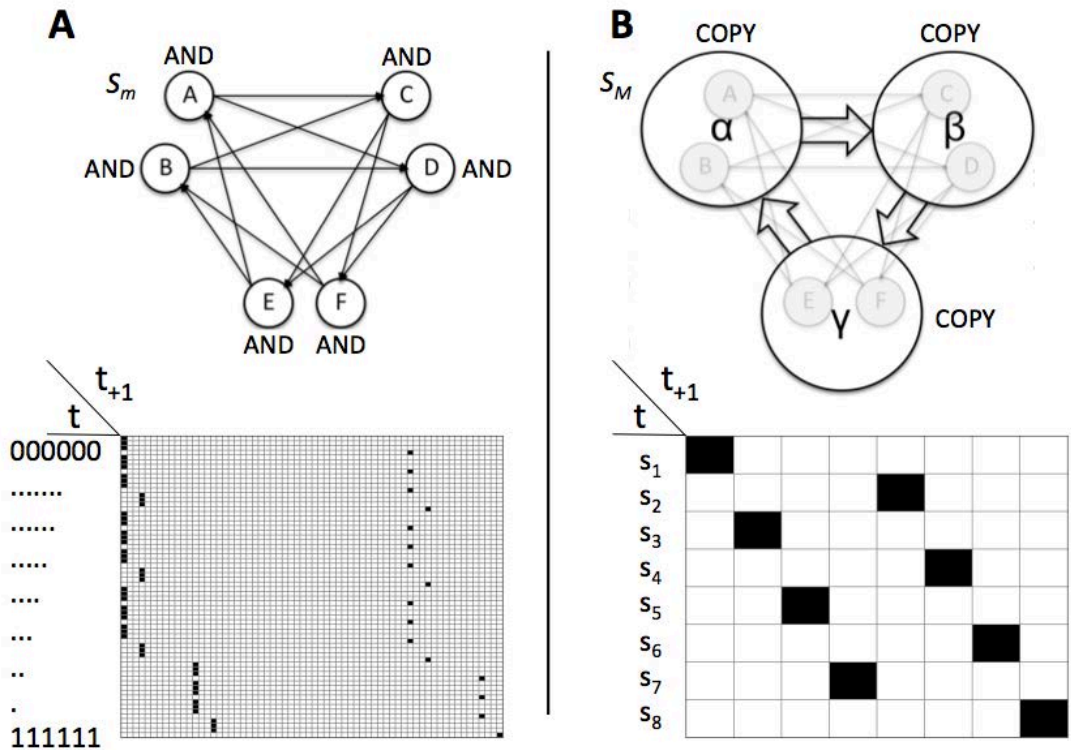

**Figure 2**. *Causal emergence via the reduction of degeneracy.* **A.** Deterministic micro causal model $S_m$ (top) and its associated TPM (bottom, in gray-scale). **B.** The macro causal model $S_M$ with $EI^{max}$, found by searching all possible macroscales, and its associated TPM (bottom).

Determinism for the system at both the microscale and macroscale is 1. But at the microscale $EI(S_m) = 2.43$ bits as $eff(S_m) = 0.41$. Effectiveness is low because degeneracy is high, at 0.59. In comparison the $S_M$ has an $EI^{max}$ value of 3 bits, as $eff(S_M) = 1$ (as degeneracy$(S_M) = 0$).

**Causal emergence as a special case of Shannon's channel capacity.** Previously, the few notions of emergence that have directly compared macro to micro have implicitly or explicitly focused solely on the concept of compression (Wolpert 2014; Shalizi and Moore 2000; Crutchfield 1994). This is understandable, given that the signature of any macro causal model is its reduced state-space. However, compression is either lossy or at best lossless. Focusing on compression ensures that the macro can at best be the compressed equivalent of the micro. In contrast, in the theory of causal emergence the dominant concept is Shannon's discovery of the capacity of a communication channel, and the ability of codes to take advantage of that to achieve reliable communication.

An information channel is composed of two finite sets, $X$ and $Y$, and a collection of transition probabilities p($y|x$) for each $x \in X$, such that for every $x$ and $y$, p($y|x$) ≥ 0 and for every $x$, $\sum_y p(y|x) = 1$ (a collection known as the channel matrix). The interpretation is that $X$ and Y are the input and output of the channel, respectively (Cover and Thomas 2012). The channel is governed by the channel matrix, which is a fixed entity. Similarly, causal structure is governed by the relationship between interventions and their effects, which is a fixed entity. Notably, both channels and causal structures can be represented as TPMs, and in a situation where a channel matrix contains the same transition probabilities as some set of state transitions, the TPMs would be identical. Causal structure is a matrix that transforms previous states into future ones.

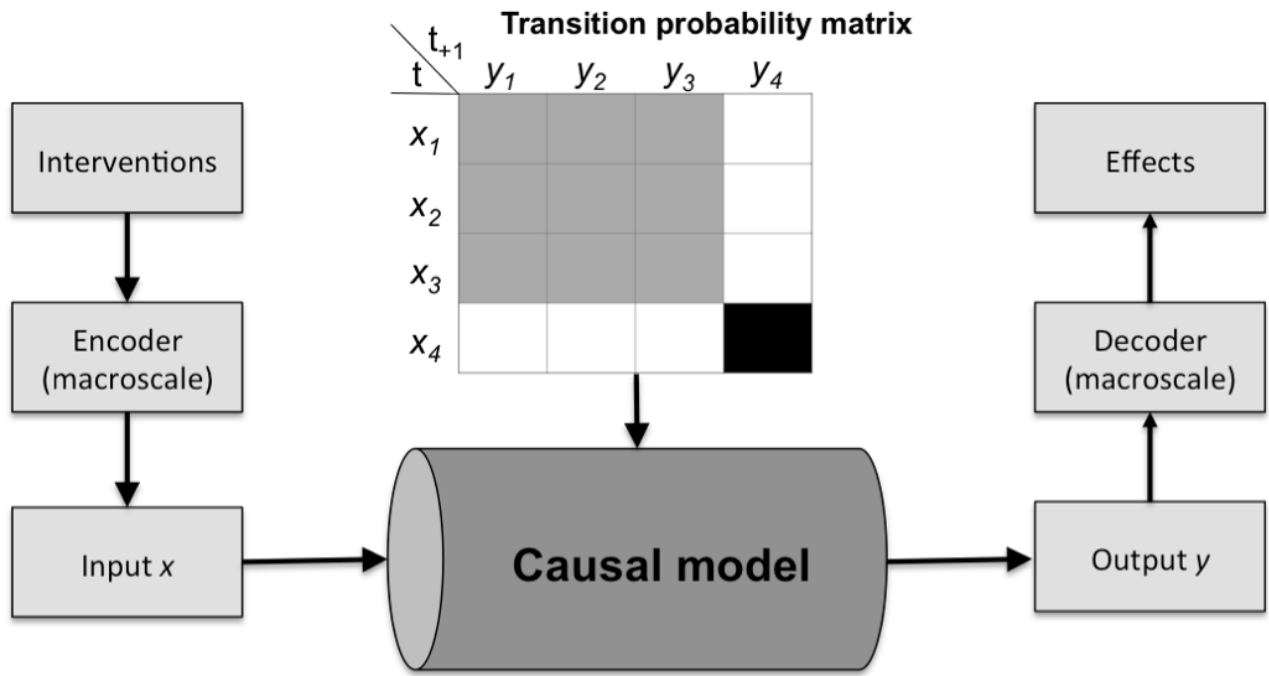

**Figure 3**. *Causal models as information channels*. Causal structure transforms interventions into effects. A macro causal model is a form of encoding for the interventions (inputs) and effects (outputs). TPM probabilities in gray scale.

Mutual information $I(X,Y)$ was originally the measure proposed by Claude Shannon (1949) to capture the rate of information that can be transmitted over a channel. Mutual information $I(X,Y)$ can be expressed as:

$$I(X,Y) = H(X) - H(X|Y)$$

which has a clear interpretation in the causal terminology already introduced. $H(X)$ represents the total possible entropy of the source, which in causal analysis is some set of interventions over all the states of the system, where all members of $p(X)$ are $1/n$, so $H(X) = \log_2(n) = size$. The conditional entropy $H(X|Y)$ captures how much information is left over about $X$ once $Y$ is taken into account. $H(X|Y)$ therefore has a clear causal interpretation as the amount of information lost by the lack of effectiveness of the system (1-*eff*). More specifically, starting with $H(X|Y) = H(X) - I(X,Y)$, which with the substitution of causal terminology is $H(X|Y) = size - (size * eff)$, shows via algebraic transposition that $H(X|Y)$ indeed captures the lack of effectiveness since $H(X|Y) = (1 - eff) * size$. Overall then:

$EI(S) = I(X,Y) = H(X) - H(X|Y) = size - ((1 - eff) * size) = size * eff$.

In comparing the macro to the micro, while $H(X)$ is necessarily decreasing at the macroscale, the conditional entropy $H(X|Y)$ may be decreasing to a greater degree, which allows the macro to beat the micro.

Shannon also proposed that communication channels have a certain capacity. The capacity is a channel's ability to transform inputs into outputs in the most informative and reliable manner. As Shannon discovered, the rate of information transmitted over a channel is sensitive to changes in the input probability distribution $p(X)$. The capacity of a channel ($C$) is found by the set of inputs that maximize the mutual information, which is also the maximal rate at which it can reliably transmit information:

$$C = max_{p(X)} I(X,Y).$$

The theory of causal emergence reveals that there is an analogous causal capacity of a system. The causal capacity ($CC$) is a system's ability to transform interventions into effects in the maximal informative and efficacious manner:

$$CC = max_{(I_D)} EI(S).$$

Just as changes to the input probability $p(X)$ to a channel can increase $I(X,Y)$, so can changes to the intervention distribution ($I_D$) in turn increase $EI$. The use of a macro causal model, and the corresponding application of macro interventions, can warp the $I_D$, leading to causal emergence. The macro causal model with $EI^{max}$ is the one that fully uses the causal capacity of the system. From this it is clear that higher spatiotemporal scales of a system are: a form of channel coding for causal structure. A macroscale of a system is a code that removes the uncertainty of causal relationships, thus using more of the available causal capacity. An example of such causal coding is given using the TPM in Figure 3 with input $X$ and output $Y$ ($t$ by $t_{+1}$):

$$X \rightarrow Y = \begin{bmatrix} 1/3 & 1/3 & 1/3 & 0 \\ 1/3 & 1/3 & 1/3 & 0 \\ 1/3 & 1/3 & 1/3 & 0 \\ 0 & 0 & 0 & 1 \end{bmatrix}.$$

To send a message through a channel matrix with these properties one defines some encoding/decoding function. The message might be some binary string like {001011010011} generated at random (such as via the application of $I_D$). The encoding function $\phi:\{message\} \rightarrow \{encoder\}$ is a rule that associates some channel input with some output, along with some decoding function $\psi$. The encoding/decoding functions together create the codebook. For simplicity issues like prefixes and instantaneous decoding are ignored here.

An encoding function is a microscale code if it is a one-to-one mapping $\phi:\{x_1, x_2, x_3, x_4\} \rightarrow \{00,01,10,11\}$ with a corresponding one-to-one decoding function:
$\psi:\{00,01,10,11\} \rightarrow \{y_1, y_2, y_3, y_4\}$, as each microstate $x$ is assumed to carry its own unique message. Assuming the message is generated by applying $I_D$ then the source has entropy $H(X) = 2$ bits as its four possible states are successive randomized interventions (so that $p(1) = 0.5$). Each code specifies a rate of transmission $R = n/t$, where $t$ is every state-transition of the system and $n$ is the number of bits sent per transition. For the microscale code of the system shown above the rate $R = 2$ bits, although these 2 bits are not sent reliably. This is because $H(X|Y)$ is large: 1.19 bits, so $I(X,Y) = H(X) - H(X|Y) = EI = 0.81$ bits. In application, this means that if one wanted to send the message {00,10,11,01,00,11}, this would take 6 state transitions (channel usages) and there would be a very high probability of numerous errors. This is because the rate exceeds the capacity at the microscale.

In contrast, we can define a macroscale encoding function as the many-to-one mapping $\Phi: \{x_1, x_2, x_3\} \rightarrow \{0\}; \{x_4\} \rightarrow \{1\}$ and similarly $\Psi: \{y_1, y_2, y_3\} \rightarrow \{0\}; \{y_4\} \rightarrow \{1\}$ such that only macrostates are assumed to carry a unique message. The rate of this code in the figure above is now twice as slow to send any message, as $R = 1$ bit, and the corresponding entropy $H(X)$ is halved (1 bit; so that $p(1) = 0.83$). However, $I(X,Y) = EI = 1$ bit, as $H(X|Y) = 0$, showing that reliable communication can proceed at the rate of 1 bit, higher than with using a microscale code. In the macroscale case, if one wanted to send the message {00,10,11,01,00,11} this would take 12 transitions (as sending {00} using this code means sending $\{x_1, x_1\}$ twice) but there would be zero errors in that transmission. While the absolute rate of channel usage has gone down, the reliability of that transmission has increased by such a degree that it outweighs the decrease. This rate of reliable communication is equal to the capacity $C$. This is identical to the necessary decrease in the *size* of the state-space of a causal model at a macroscale, and how this can be outweighed by an increase in the *eff*. One can think of the channel efficiency as the coefficient $I(X,Y)/H(X)$ (Press et al. 2007), and the search across input distributions increases this coefficient. This is just how the effectiveness of a causal structure is $EI/size$, and the search across models increases the effectiveness.

Interestingly, it is provable causal emergence requires symmetry breaking. A channel is defined as symmetric when its rows $p(y|x)$ and columns are permutations of each other. A channel is weakly symmetric if the row probabilities are permutations of each other and all the column sums are equal. For such a symmetric channel the input distribution that generates $I^{max}$ has been proven to be the uniform distribution $H^{max}$ (Cover and Thomas 2012). When using the micro causal model of a system, $I_D = H^{max}$. Therefore, for symmetric or weakly symmetric systems, the microscale provides the best causal model without any need to search across model space. It is only in systems with asymmetrical causal relationships that causal emergence can occur.

**Causal capacity can approximate channel capacity as model choice increases.** The causal model that uses the full causal capacity of a system has an associated $I_D$, which achieves its success in the same manner as the input distribution that uses the full channel capacity: by sending only a subset of the possible messages during channel usage. But while the causal capacity is bounded by the channel capacity, it is not always identical to it. Because the warping of $I_D$ is a function of model choice, which is constrained in various ways (a subset of possible distributions), causal capacity is a special case of the more general channel capacity (defined over all possible distributions). Coarse-graining is one way to manipulate (warp) $I_D$: by moving up to a macro scale. It is not the only way. Choices made in causal modeling, including the choice of scale, but also choice of initial condition, and whether to classify variables as exogenous or endogenous to the model, can all also warp $I_D$.

For example, consider a system that's a Markov chain of 8 states:

$$S_m = \begin{bmatrix} 1/8 & 1/8 & 1/8 & 1/8 & 1/8 & 1/8 & 1/8 & 1/8 \\ 1/8 & 1/8 & 1/8 & 1/8 & 1/8 & 1/8 & 1/8 & 1/8 \\ 1/8 & 1/8 & 1/8 & 1/8 & 1/8 & 1/8 & 1/8 & 1/8 \\ 1/8 & 1/8 & 1/8 & 1/8 & 1/8 & 1/8 & 1/8 & 1/8 \\ 1/8 & 1/8 & 1/8 & 1/8 & 1/8 & 1/8 & 1/8 & 1/8 \\ 1/8 & 1/8 & 1/8 & 1/8 & 1/8 & 1/8 & 1/8 & 1/8 \\ 0 & 0 & 0 & 0 & 0 & 0 & 1 & 0 \\ 0 & 0 & 0 & 0 & 0 & 0 & 0 & 1 \end{bmatrix}$$

where some states in the chain are completely random in their transitions. If the $I_D$ contains all possible states, $EI = 0.63$ bits. Yet every causal model implicitly classifies variables as endogenous or exogenous to the model. For instance, here, we can take only the last two states ($s_7$, $s_8$) as endogenous to the macro causal model, while leaving the rest of the states as exogenous. This restriction is still a macro model because it has a smaller state-space. For this macro causal model of the system $EI = 1$ bit, meaning that causal emergence occurs, again because the $I_D$ is warped by model choice. This warping can itself be quantified as the loss of entropy in the intervention distribution, $H(I_D)$.

$$I_D(warped) = [0 \quad 0 \quad 0 \quad 0 \quad 0 \quad 0 \quad 1/2 \quad 1/2]$$
$$\uparrow$$
$$I_D = [1/8 \quad 1/8 \quad 1/8 \quad 1/8 \quad 1/8 \quad 1/8 \quad 1/8 \quad 1/8]$$

While leaving noisy or degenerate states exogenous to a causal model can lead to causal emergence, so can leaving certain elements exogenous.

Notably, there are multiple ways to leave elements exogenous (to not explicitly include them in the macro model). For instance, exogenous elements are often implicit background assumptions in causal models. Such background conditions may consist of setting an exogenous element to a particular state (freezing) for the causal analysis, or setting the system to an initial condition. Alternatively, one could allow an element to vary under the influence of the applied $I_D$. This latter form has been called "black boxing", where an element's internal workings, or role in the system, cannot be examined (Ashby 1956; Bung 1963). In Figure 4, both types of model choices are shown, each leading to causal emergence.

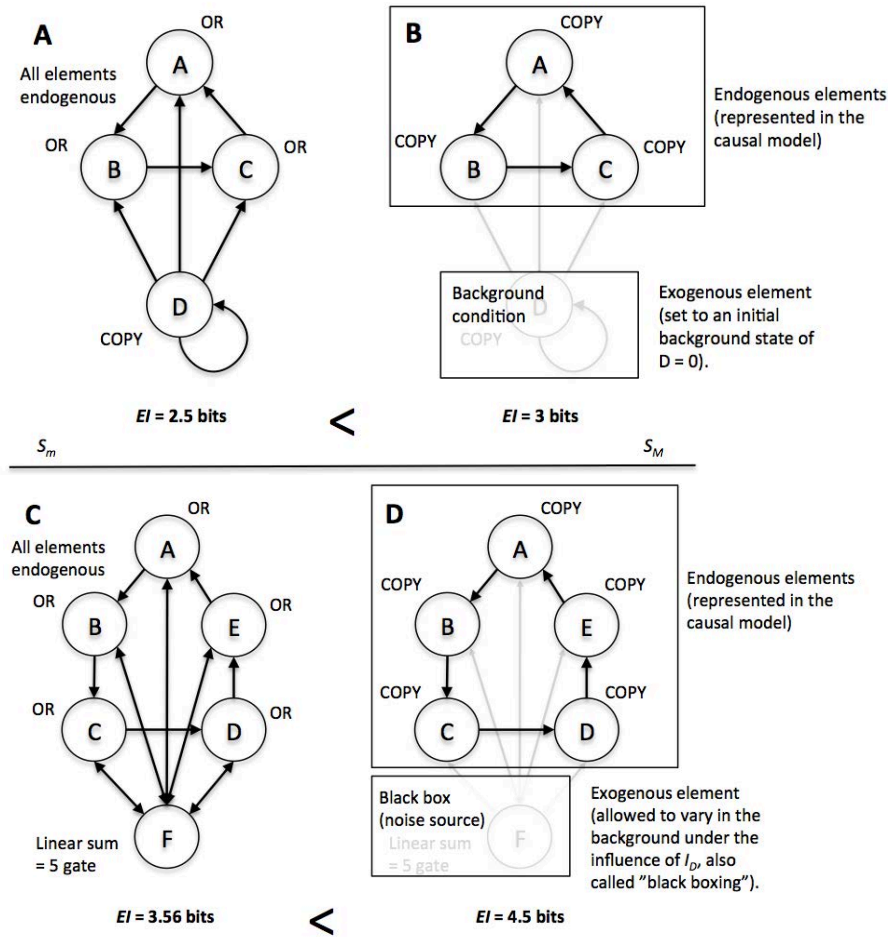

**Figure 4**. *Multiple types of model choice can lead to causal emergence.* **A.** The full microscopic model of the system, where all elements are endogenous. **B.** The same system but modeled at a macroscale where only elements {ABC} are endogenous, while {D} is exogenous: it was set to an initial state of 0 as a background condition of the causal analysis. **C.** The full microscopic model of a system with six elements. **D.** The same system as in C but at a macroscale with the element {F} exogenous: it varies in the background in response to the application of the $I_D$. *EI* is higher for both macro causal models.

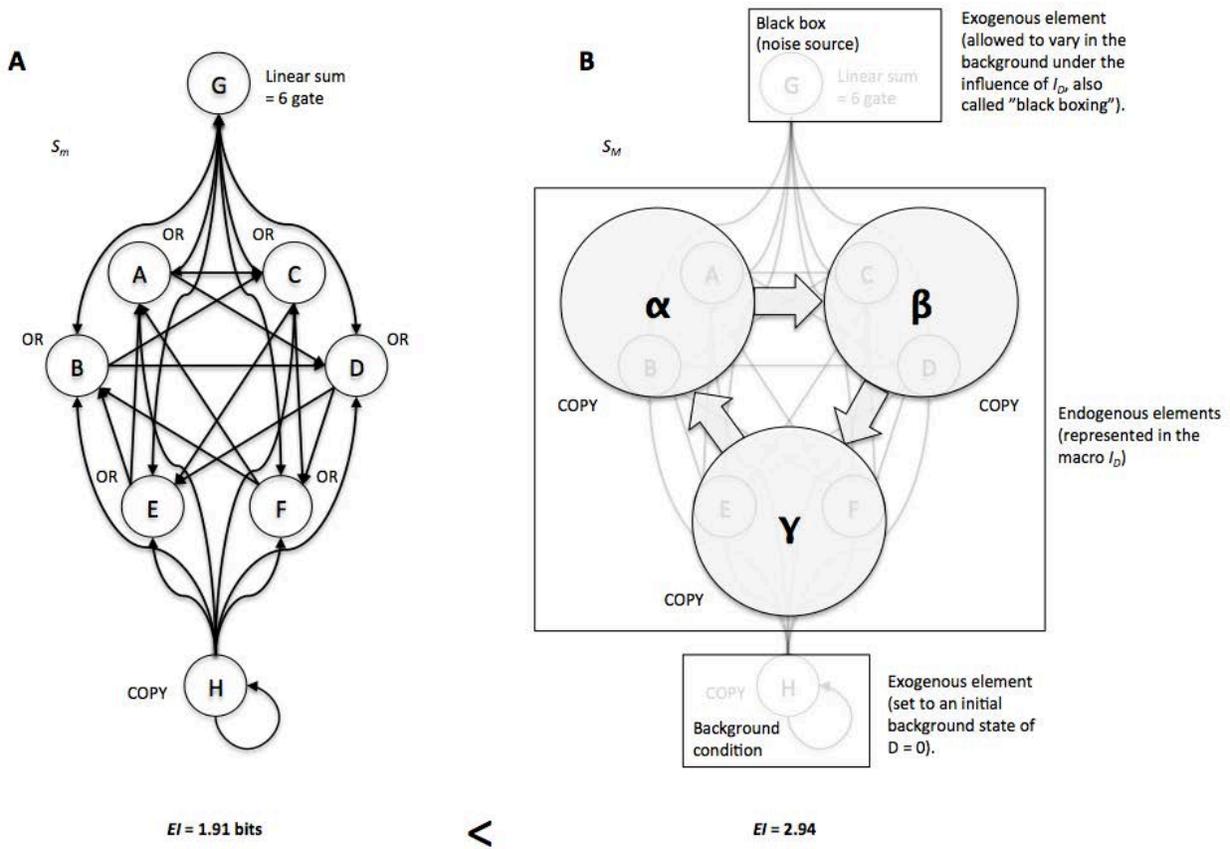

**Figure 5**. *A macro model that demonstrates different types of model choice.* **A.** The microscale of an eight-element system. **B.** The same system with some elements coarse-grained, others "black boxed", and some left in a particular initial condition.

Rather than dictating precisely what types of model choices are "allowable" in the construction of causal models, a more general principle can be distinguished: the more ways to warp $I_D$ via model-building choices, the closer the causal capacity approximates the actual channel capacity. For example, consider the system in Figure 5A. In Fig. 5B, a macroscale is shown that demonstrates causal emergence using various types of model choice (by coarse-graining, black-boxing an element, and setting a particular initial condition for an exogenous element). As can be seen in Fig. 6, the more degrees of freedom in terms of model choice there are, the closer the causal capacity approximates the channel capacity. The channel capacity of this system was found via gradient ascent after the simulation of millions of random probability distributions $p(X)$, searching for the one that maximizes $I$.

Model choice warps the microscale $I_D$ in such a way that it moves closer to $p(X)$, as shown in Fig. 6B. As model choice increases, the $EI^{max}$ approaches the $I^{max}$ of the channel capacity (Fig. 6C).

Causal emergence is a phenomenon based on using information theory to quantify and capture causation and causal structure. What of measures beyond $EI$? Indeed, it has already been proven that integrated information, which has similarities to $EI$ but is a more complex measure, can peak at both coarse-grained macroscales (Hoel et al. 2016) and "black boxed" macroscales (Marshall et al. 2016). One consequence of this research is that the more complex the informational measure of causation, the more possible model choices there will be that will take advantage of the causal capacity of a system.

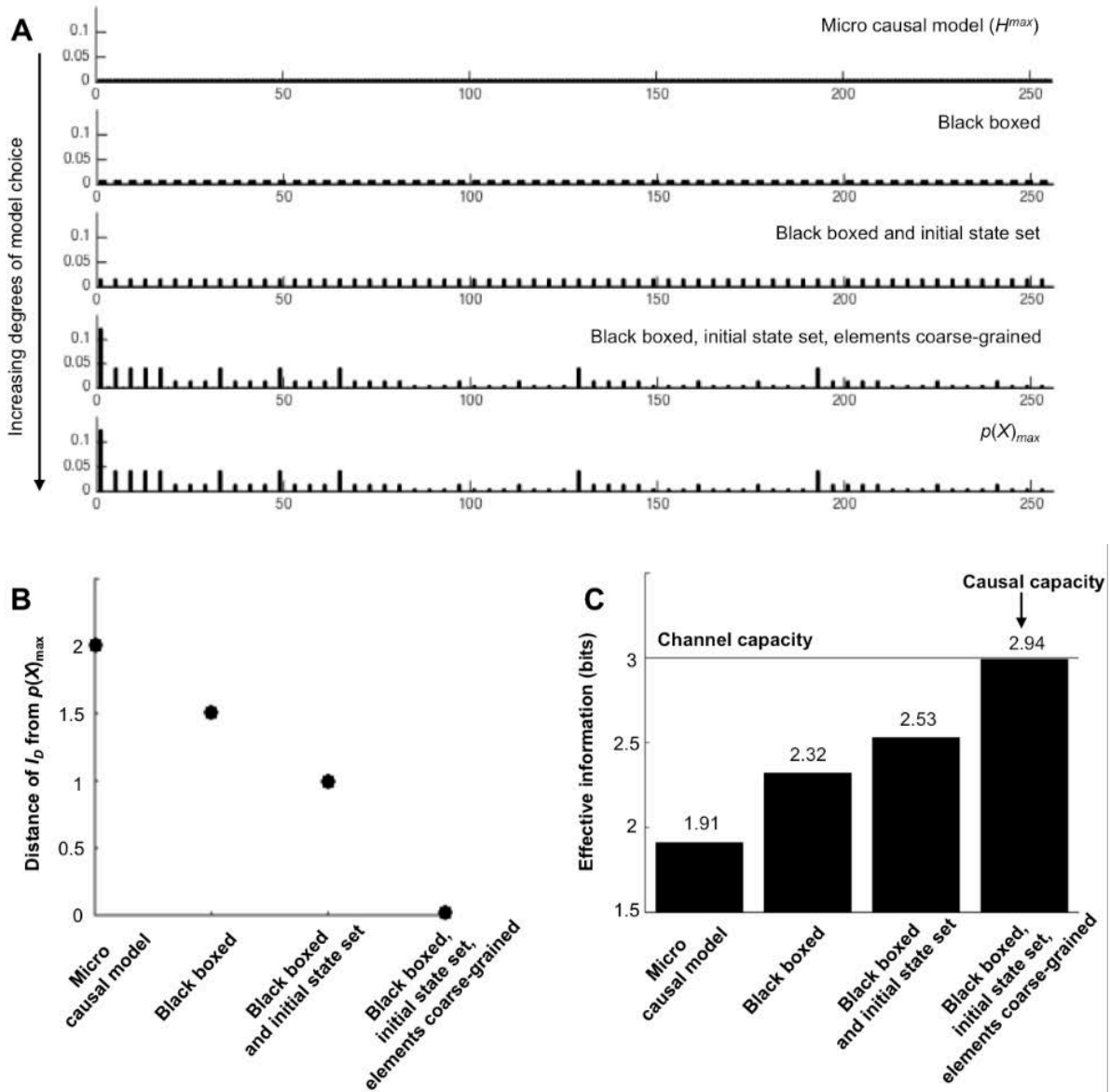

**Figure 6**. *Causal capacity approximates the channel capacity as more degrees of freedom in model choices are allowed.* **A.** The $I_D$ of various models for the system in Fig.5, getting closer to the $p(X)$ that gives $I^{max}$. **B.** The earth mover's distance (Rubner et al. 2000) from each $I_D$ to the input distribution $p(X)$ of the channel capacity $I^{max}$, as more model set is added. **C.** Increasing degrees of model choice leads to an $EI^{max}$ that is close to $I^{max}$ (all these steps together end in the macroscale shown in Fig. 5B).

**Discussion.** The theory of causal emergence directly challenges the reductionist assumption that the best causal model of any system is its microscale. Causal emergence reveals a contrasting and counterintuitive phenomenon: sometimes the map is better than the territory. As shown here, this has a precedent with Shannon's discovery of the capacity of a communication channel, which was also thought of as counterintuitive at the time. Here an analogous idea of a causal capacity of a system has been developed using effective information. Effective information, an information-theoretic measure of causation, is assessed by the application of an intervention distribution. However, in the construction of causal models, model choice warps the intervention distribution, which in turn reveals a system's causal capacity, which might be much greater than that used by its microscale representation. Causal capacity can approach the channel capacity, particularly as degrees of model choice increase. The choice of a model's scale, its elements and states, its background conditions, its initial conditions, what variables to leave exogenous and in what manner, and so on, all result in warping of the intervention distribution $I_D$. All of these make the state-space of the system smaller, so can be classified as macroscales, yet all may possibly lead to causal emergence.

Previously, some have argued (Campbell 1974; Ellis 2016) that macroscales may enact top-down causation: that the higher-level laws of a system somehow constrain or cause in some unknown way the lower-law levels. Top-down causation has been considered as contextual effects, such as a wheel rolling downhill (Sperry 1969), or as higher scales fulfilling the different Aristotelian types of causation (Auletta et al. 2008; Ellis 2015). However, varying notions of causation, as well as a reliance on examples where the different scales of a system are left loosely defined, has led to much confusion between actual causal emergence (when higher scales are truly causally efficacious) and things like contextual effects, whole-part relations, or even mental causation, all of which are lumped together under "top-down causation." In comparison, the theory of causal emergence can rigorously prove that macroscales are error-correcting causal codes, and that many systems have a causal capacity that exceeds their microscale representations. Note that the theory of causal emergence does not contradict other theories of emergence, such as proposals that truly novel laws or properties may come into being at higher scales in systems (Broad 1925). It does, however, indicate that for some systems only modeling at a macroscale uses the full causal capacity of the system; it is this way that higher scales can have causal influence above and beyond the microscale.

One objection to causal emergence is that, since it requires asymmetry, it is trivially impossible if the microscale of a system happens to be composed only of causal relationships in the form of logical biconditionals (for which effectiveness = 1, as determinism = 1 and degeneracy = 0). This is a stringent condition, yet whether or not this is true for physical systems I take as an open question, as deriving causal structure from physics is an unfinished research program (Frisch 2014), as is physics itself. Even if it were true that the microscale of physics has this stringent biconditional causal property for all systems, the theory of causal emergence would still show us how to pick scales of interest in systems where the microscale is unknown,, difficult to model, or subject to noise at the lowest level we can experimentally observe or intervene upon.

Another possible objection to causal emergence is that it is not natural but rather enforced upon a system via an experimenter's application of an intervention distribution. For formalization purposes, it is the applier of the $do(x)$ operator who is the source of the intervention distribution. However, nature itself may intervene upon a system with statistical regularities, just like an intervention distribution. Some of these naturally occurring input distributions may have a viable interpretation as a macro causal model. In this sense, some systems may be naturally microscale or macroscale, depending on their own causal capacity and the probability distribution of some natural source of driving input.

The application of the theory of causal emergence to fields such as neuroscience is already underway (Hoel 2016). Specifically, it may help solve longstanding problems in neuroscience involving scale, such as the debate over whether brain circuitry functions at the scale of neural ensembles or individual neurons (Buxhoeveden and Casanova 2002; Yuste 2015). It has also been proposed that the brain integrates information at a higher level (Tononi 2008) and it was proven that integrated information can indeed peak at a macroscale in Hoel et al. (2016). An experimental way to resolve these debates is to systematically measure $EI$ in ever-larger groups of neurons, eventually arriving at or approximating the cortical scale with $EI^{max}$ (Hoel et al. 2013; Tononi et al. 2016). If there are such privileged scales in a system then intervention and experimentation should focus on those scales.

Finally, the phenomenon of causal emergence provides a general explanation as to why science and engineering take diverse spatiotemporal scales of analysis, regularly consider systems as isolated from other systems, and only consider a small repertoire of physical states in particular initial conditions. It is because scientists and engineers unwittingly search for causal models that use a significant portion of a system's causal capacity, rather than just building the most detailed microscopic causal model possible.

## Acknowledgements

Thanks to Giulio Tononi, Larissa Albantakis, and Billy Marshall for all their support during my PhD.

## Citations

Ashby, William Ross. "An introduction to cybernetics." *An introduction to cybernetics (1956)*.

Auletta, Gennaro, G. F. R. Ellis, and Luc Jaeger. "Top-down causation by information control: from a philosophical problem to a scientific research programme." *Journal of The Royal Society Interface* 5.27 (2008): 1159-1172.

Ay, Nihat, and Daniel Polani. "Information flows in causal networks." *Advances in complex systems* 11.01 (2008): 17-41.

Broad, Charlie Dunbar. *The mind and its place in nature*. Routledge, 2014 (first pub: 1925).

Bunge, Mario. "A general black box theory." *Philosophy of Science* (1963): 346-358.

Buxhoeveden, Daniel P., and Manuel F. Casanova. "The minicolumn hypothesis in neuroscience." *Brain* 125.5 (2002): 935-951.

Frisch, Mathias. *Causal reasoning in physics*. Cambridge University Press, 2014.

Campbell, Donald T. "'Downward causation'in hierarchically organised biological systems." *Studies in the Philosophy of Biology*. Macmillan Education UK, 1974. 179-186.

Cover, Thomas M., and Joy A. Thomas. *Elements of information theory*. John Wiley & Sons, 2012.

Crutchfield, James P. "The calculi of emergence: computation, dynamics and induction." *Physica D: Nonlinear Phenomena* 75.1 (1994): 11-54.

Davidson, Donald. *Essays on actions and events: Philosophical essays*. Vol. 1. Oxford University Press on Demand, 2001. (first pub: 1970)

Ellis, George. *How can physics underlie the mind?* Springer, 2016.

Ellis, George. "Recognising top-down causation." *Questioning the Foundations of Physics*. Springer International Publishing, 2015. 17-44.

Fisher, R. A. (1935). The design of experiments.

Fodor, J. A. (1974). Special sciences (or: the disunity of science as a working hypothesis). *Synthese*, *28*(2), 97-115.

Granger, Clive WJ. "Investigating causal relations by econometric models and cross-spectral methods." *Econometrica: Journal of the Econometric Society* (1969): 424-438.

Hoel, Erik P. *Brain organization and information integration*. PhD thesis at the University of Wisconsin-Madison (2016).

Hoel, Erik P., et al. "Can the macro beat the micro? Integrated information across spatiotemporal scales." *Neuroscience of Consciousness* 2016.1 (2016).

Hoel, Erik P., Larissa Albantakis, and Giulio Tononi. "Quantifying causal emergence shows that macro can beat micro." *Proceedings of the National Academy of Sciences* 110.49 (2013): 19790-19795.

Janzing, D., Balduzzi, D., Grosse-Wentrup, M., & Schölkopf, B. (2013). Quantifying causal influences. *The Annals of Statistics*, *41*(5), 2324-2358.

Korb, Kevin B., Lucas R. Hope, and Erik P. Nyberg. "Information-theoretic causal power." *Information Theory and Statistical Learning*. Springer US, 2009. 231-265.

Kullback, Solomon. *Information theory and statistics*. Courier Corporation, 1997.

Kim, Jaegwon. *Mind in a physical world: An essay on the mind-body problem and mental causation*. MIT press, 2000.

Marshall, W., Albantakis, L., & Tononi, G. (2016). Black-boxing and cause-effect power. *arXiv preprint arXiv:1608.03461*.

Massey, James. "Causality, feedback and directed information." *Proc. Int. Symp. Inf. Theory Applic.(ISITA-90)*. 1990.

Press, W. H., et al. "Conditional entropy and mutual information." *Numerical Recipes 3rd Edition: The Art of Scientific Computing, Cambridge University Press, New York* (2007).

Rubner, Y., Tomasi, C., & Guibas, L. J. (2000). The earth mover's distance as a metric for image retrieval. *International journal of computer vision*, *40*(2), 99-121.

Schreiber, Thomas. "Measuring information transfer." *Physical review letters* 85.2 (2000): 461.
Shalizi, Cosma Rohilla, and Cristopher Moore. "What is a macrostate? Subjective observations and objective dynamics." *arXiv preprint cond-mat/0303625* (2003).

Shannon, Claude E. "Communication theory of secrecy systems." *Bell system technical journal* 28.4 (1949): 656-715.

Sperry, Roger W. "A modified concept of consciousness." *Psychological review* 76.6 (1969): 532.

Stalnaker, Robert. "Varieties of supervenience." *Philosophical perspectives* 10 (1996): 221-241.

Tononi, Giulio, et al. "Integrated information theory: from consciousness to its physical substrate." *Nature Reviews Neuroscience* (2016).

Tononi, Giulio. "Consciousness as integrated information: a provisional manifesto." *The Biological Bulletin* 215.3 (2008): 216-242.

Tononi, Giulio, and Olaf Sporns. "Measuring information integration." *BMC neuroscience* 4.1 (2003): 1.

Tononi, Giulio. "Information measures for conscious experience." *Archives italiennes de biologie* 139.4 (2001): 367.

Wolpert, David H., et al. "Optimal high-level descriptions of dynamical systems." *arXiv preprint arXiv:1409.7403* (2014).

Yuste, Rafael. "From the neuron doctrine to neural networks." *Nature Reviews Neuroscience* 16.8 (2015): 487-497.